\documentstyle[11pt,epsfig]{article}

\renewcommand\({\left(}
\renewcommand\){\right)}
\renewcommand\[{\left[}
\renewcommand\]{\right]}

\newcommand{\ra}{\rightarrow}
\def\lsim{\raise 0.4ex\hbox{$<$}\kern -0.8em\lower 0.62
ex\hbox{$\sim$}}
\def\gsim{\raise 0.4ex\hbox{$>$}\kern -0.7em\lower 0.62
ex\hbox{$\sim$}}

\newcommand{\hogw}{h_0^2\Omega_{\rm gw}}

\newcommand\eq[1]{eq.~(\ref{#1})}
\newcommand\eqs[2]{eqs.~(\ref{#1}) and (\ref{#2})}

\newcommand\ee{\end{equation}}
\newcommand\be{\begin{equation}}
\newcommand\ees{\end{eqnarray}}
\newcommand\bees{\begin{eqnarray}}
\newcommand\sub[1]{_{\rm #1}}

\def\dslash{\not{\hbox{\kern-2pt $\partial$}}}
\def\Dslash{\not{\hbox{\kern-4pt $D$}}}
\def\pslash{\not{\hbox{\kern-2.3pt $p$}}}
\newcommand{\bea}{\begin{eqnarray*}}
\newcommand{\eea}{\end{eqnarray*}}
\newcommand{\bean}{\begin{eqnarray}}
\newcommand{\eean}{\end{eqnarray}}

\newcommand{\half}{ \frac {_1}{^2} }

\newcommand{\const}{ {\rm const} }

\newcommand{\uno}{ 1 \!\!\!\! \:\: {\rm l}\!\!\!\!\!\! \:\:\: \_ }
\newcommand{\s}{ {\cal {S}} }

\newcommand{\lag}{{\cal {L}}}
\newcommand{\di}{ {\rm d} }
\newcommand{\de}{\partial}

\newcommand{\obd}{ \omega _{\rm BD} }
\newcommand{\gmn}{ g _{\mu \nu} }
\newcommand{\Gmn}{ g ^{\mu \nu} }
\newcommand{\etamn}{\eta _{\mu \nu}}
\newcommand{\Etamn}{\eta ^{\mu \nu}}
\newcommand{\hmn}{h _{\mu \nu}}

\newcommand{\thetamn}{\theta _{\mu \nu}}
\newcommand{\Thetamn}{\theta ^{\mu \nu}}
\newcommand{\epsm}{\epsilon _\mu}
\newcommand{\Epsm}{\epsilon ^\mu}
\newcommand{\eplmn}{ e ^{ (+) } _{ \mu \nu} }
\newcommand{\Aplus}{ A^{ (+) } }
\newcommand{\ecrmn}{ e ^{( \times)} _{ \mu \nu} }
\newcommand{\Across}{ A^{ (\times) } }
\newcommand{\esmn}{ e ^{\rm (s) } _{ \mu \nu} }

\newcommand{\n}{{\bf \hat n} }
\newcommand{\m}{{\bf \hat m} }
\newcommand{\x}{{\bf \hat x} }
\newcommand{\y}{{\bf \hat y} }

\newcommand{\uarm}{{\bf \hat u} }
\newcommand{\varm}{{\bf \hat v} }
\newcommand{\dOn}{\di \Omega _\n}

\newcommand{\duno}{ D^{ (1) } }
\newcommand{\ddue}{ D^{ (2) } }
\newcommand{\G}{\Gamma _{ijkl} }
\newcommand{\A}{ {\cal{A}} _{ijkl} }
\newcommand{\B}{ {\cal{B}} _{ijkl} }
\newcommand{\C}{ {\cal{C}} _{ijkl} }
\newcommand{\D}{ {\cal{D}} _{ijkl} }
\newcommand{\E}{ {\cal{E}} _{ijkl} }

\newcommand{\tr}{ {\rm Tr} }

\newcommand{\jzero}{ j_0 (\alpha) }
\newcommand{\juno}{ j_1 (\alpha) }
\newcommand{\jdue}{ j_2 (\alpha) }

\newcommand{\las}{ _{\rm laser} }

\newcommand{\R}{ {\cal {R}} }

\newcommand{\disp}[1]{ \displaystyle{#1} }

\begin{document}

\begin{titlepage}
\begin{flushright}
IFUP-TH 40/99\\
July 1999\\
\end{flushright}
\vspace{5mm}
\begin{center}
{\Large \bf 
Detection strategies for  scalar gravitational\\

\vspace{4mm}
waves with interferometers and resonant spheres\\}
\vspace{.4in}
{\large\bf  Michele Maggiore}$^{a}$ and 
{\large\bf  Alberto Nicolis}$^{a,b}$\\
\vspace{.6 cm}
{\em (a) INFN, sezione di Pisa, and Dipartimento di Fisica, Universit\`a
  di Pisa\\
via Buonarroti 2, I-56127 Pisa, Italy\\}
{\em (b) Scuola Normale Superiore, Piazza dei Cavalieri 7, I-56125
Pisa.}\\
\end{center}

\vspace{.6cm}
\begin{abstract}
\noindent
We compute the response and the  angular pattern function 
of an interferometer  for a  scalar component of
gravitational radiation in Brans-Dicke theory. We examine the problem
of detecting a stochastic background of scalar GWs and compute
the scalar overlap reduction function  in the correlation
between an interferometer and the monopole mode of a resonant sphere.
While the correlation between two interferometers is maximized taking
them as close as possible, the interferometer-sphere correlation is
maximized at a finite value of $f\times d$, where $f$ is the resonance
frequency of the sphere and $d$ the distance between the detectors.
This defines an 
optimal resonance frequency of the sphere as a function of the
distance. For the correlation between the Virgo interferometer
located near Pisa and a
sphere located in Frascati, near Rome, 
we find an optimal resonance frequency 
$f\simeq 590$ Hz. We also briefly discuss the difficulties in applying this
analysis to the dilaton and moduli fields predicted by string theory.
\end{abstract}

\vspace{.6cm}

\end{titlepage}

\newpage

\section{Introduction}
A number of 
interferometers for gravitational wave (GW) detection are presently 
under constructions and are expected to be operating in the next
few years. In particular, 
VIRGO is being built near Pisa, the two LIGO interferometers
are being built in the US, GEO600 near Hannover, and TAMA300 in Japan.
These interferometers are in principle sensitive also to a
hypothetical scalar component of gravitational
radiation.  Scalar GWs appear already in the
simplest generalization of General Relativity, namely Brans-Dicke
theory, whose action reads
\be\label{BD}
\s_{\rm BD}=\frac{1}{16\pi}\int d^4x\sqrt{-g}\,\[
\varphi R-\frac{\obd}{\varphi}\nabla^{\mu}\varphi\nabla_{\mu}\varphi\]
+\s\sub{matter}
\, ,
\ee
with $\varphi$ the Brans-Dicke scalar. The coupling of matter with
gravity, $\s \sub{matter}$, is dictated by the equivalence principle. 
In order to avoid conflict with solar system experiments, one must
take $|\obd |$ greater than approximately 600. 

At a more fundamental
level, various scalar fields with 
interactions of gravitational strength
come from string theory. A universal example is the
dilaton $\Phi$; 
the low energy effective action of
string theory, in the graviton-dilaton sector, reduces to 
the first term on the right-hand side of \eq{BD}, with
the identifications $\varphi =(16\pi/\alpha ')e^{-2\Phi}$ 
(where $\alpha '$ is the string tension) and
$\obd =-1$. To avoid conflict with experiments, it is expected that
the dilaton will get a mass from non-perturbative
mechanisms~\cite{TV}, or that it decouples from matter with the
cosmological mechanism proposed in~\cite{DP}. 
Furthermore, various  scalar fields (moduli) 
appear when
compactifying string theory from ten to four dimensions. Their number 
and couplings are
strongly dependent on the specific compactification used.

In this paper we investigate whether it is possible to search for
such scalar particles using the GW interferometers under construction, 
as well as the resonant spheres which are under study. 
We start from the Brans-Dicke theory and,
in sect.~2,
we  discuss the response of an interferometer to a GW with a scalar
component: in particular, we find that such a scalar component creates a
transverse (with respect to the direction of propagation of the GW) stress in the
detector; we compute the phase shift $\Delta \varphi$ measured in the interferometer
and derive the angular pattern function, i.e. the dependence of the signal on
the direction $(\theta, \phi)$ of the impinging GW (see figure~\ref{terna1}).
We find $\Delta \varphi \propto \sin ^2 \theta \cos 2\phi$.
We also show the physical (and formal) equivalence of two different gauges
used to describe scalar radiation. 

In sect.~3 we consider the
detection of a stochastic background of scalar GWs. In this case it is
necessary to correlate two different detectors. We 
give a general treatment of the computation of the overlap reduction 
function $\Gamma (f)$
for generic detectors in the scalar case: such a function represents a measure
of the correlation between the signals of the two detectors and depends
on the frequency of observation $f$ and on the type of detectors one
uses, as well as on their
location and relative orientation. Similarly to what has been
done in ref.~\cite{Fla} in the case
of the $+$ and $\times$ components, we ``factorize out'' from 
$\Gamma(f)$ the response 
tensors $D_{ij}$ of the detectors, which summarize the whole
information 
about the type of
the detectors and their orientation in space; next we compute explicitly the
remaining part of $\Gamma(f)$, that is the dependence on the frequency
and 
the location
of the detectors. The result is thus completely general and it is 
applicable to any given pair of detectors.

We then examine in particular the
correlation between the VIRGO interferometer and a resonant sphere,
as the prototype which is
presently under study in Frascati, near
Rome~\cite{BCCFF,SPHERA,BBCFL,BCFF,Bru}. Similar correlations are also
in principle possible between LIGO and the TIGA resonant sphere located
in Louisiana~\cite{MJ}.
We compute the interferometer-sphere 
overlap reduction function and we find that, contrarily to what
happens in the correlation of two interferometers, the correlation is
not optimized when the detectors are as close as possible (compatibly
with the constraint of decorrelating local noises), but instead
there is an optimum non zero
value of the product of the distance between the detectors and  the
resonance frequency of the sphere. 
For the distance between  the VIRGO interferometer in Cascina, near
Pisa, and
a resonant sphere in Frascati, near Rome, we find an optimum correlation if
the resonance frequency of the sphere is $f\simeq 590$ Hz. This value
of $f$ is quite interesting because is in the range where
interferometers achieve their highest sensitivities and at the same time
is comparable to realistic values for the resonance
frequency of the spheres which are presently under study. 
Actually, the TIGA prototype has its first resonant mode at
3.2 kHz, but hollow spheres, which are presently at the stage of
preliminary feasibility studies~\cite{CFFLO}, 
depending on the material used and other
parameters, could have a
resonance frequency  between 200 Hz and 1-2 kHz, with a
bandwidth of order 20 Hz.

In sect.~4 we discuss how the interaction with the detectors is modified by a very 
small mass term for the scalar field. This is partly motivated  by the desire to examine the
perspective for detection of the string dilaton and moduli. We find that, in presence of
such a mass term, the stress induced in the detector by the scalar
wave is 
not anymore purely
transverse, but has also a longitudinal component of relative amplitude $m^2 / \omega^2$,
if $m$ is the mass of the scalar and $\omega$ is the frequency of the GW. 

In sect.~5 we will then  briefly discuss some difficulties in applying our analysis 
to these fields predicted by string theory. A number of technical details are 
collected in the appendixes.

\section{The response of the interferometer to scalar gravitational waves}

\subsection{Computation in the gauge $\esmn ={\rm diag}(0,1,1,0)$}
In this section we compute the phase shift measured in the
interferometer when a scalar GW is coming from an arbitrary
direction. There are of course
different possible gauge choices (i.e. coordinate transformations)
for representing plane wave solutions
of the equations of motion of Brans-Dicke theory
with both spin 2 and spin 0 components. We first
consider a gauge choice that, for a wave propagating in the $+z$
direction,  brings the metric perturbation  in the form
\be\label{h_munu}
\hmn (t-z) = \Aplus (t-z) \: \eplmn + \Across (t-z) \: \ecrmn 
+ \Phi (t-z) \: \esmn   \; ,
\ee
where $e_{\mu\nu}^{+,\times , \,{\rm s}}$ are the  polarization tensors,
\be\label{emn}
\eplmn          \equiv          \left(  \begin{array}{cccc}
                                0 & 0 & 0 & 0 \\
                                0 & +1 & 0 & 0 \\
                                0 & 0 & -1 & 0 \\
                                0 & 0 & 0 & 0
                                \end{array}     \right) \, ,
\hspace{10mm}
\ecrmn          \equiv          \left(  \begin{array}{cccc}
                                0 & 0 & 0 & 0 \\
                                0 & 0 & +1 & 0 \\
                                0 & +1 & 0 & 0 \\
                                0 & 0 & 0 & 0
                                \end{array}     \right)\, ,
\ee
\be\label{esmn}
\esmn           \equiv          \left(  \begin{array}{cccc}
                                0 & 0 & 0 & 0 \\
                                0 & +1 & 0 & 0 \\
                                0 & 0 & +1 & 0 \\
                                0 & 0 & 0 & 0
                                \end{array}     \right)  \; . 
\ee
The choice of gauge that brings the plane wave
solutions of the equation of motion
of Brans-Dicke theory 
into this form
is discussed in ref.~\cite{Lee,BBCFL} and, for
comparison with later results, 
we recall the main points of the derivation in app.~A.
Under rotations around the $z$ axis, it is straightforward to verify that
$e^{(+)} _{ij} \pm i \: e^{(\times)} _{ij}$ have helicities $\pm2$, while
$e^{\rm (s)} _{ij}$ has helicity 0. Thus
the term $A ^{(+)}\: \eplmn + A ^{(\times)} \: \ecrmn$
describes  ordinary gravitational waves 
with $+$ and $\times$  polarizations 
in the  transverse  traceless gauge, and
the term $\Phi \: \esmn$, describe a scalar GW,
characteristic of the  theory that we are considering.

To compute the response of the interferometer we start with 
the geodesic equation for a free-falling mass,
\be             \label{geodetiche1}
\ddot x^\mu + \Gamma^\mu_{\alpha \beta} 
\: \dot x^\alpha \: \dot x^\beta = 0   \; ,
\ee
where
\be
\Gamma^\mu_{\alpha \beta} = \half \Etamn \left( 
\de_\alpha h_{\beta \nu} + \de_\beta h_{\alpha \nu} 
- \de_\nu h_{\alpha \beta}
\right) 
\ee
are the linearized Christoffel symbols 
and $\dot {(\:\:)} \equiv \disp{\frac{\di}{\di\tau}}$
denotes derivation with respect to proper-time of the mass.
In the gauge (\ref{h_munu}) we have  $\Gamma ^\mu_{00}=0$ and therefore,
if the mass is initially at rest, it will remain at rest also
when the wave arrives, and the proper time $\tau$ is
the same as the time variable $t$.
In other words,  choosing the
gauge (\ref{h_munu}) means that  we automatically choose the reference frame
 whose coordinates are comoving
with free-falling masses.

The situation is then perfectly analogous to the ordinary
gravitational waves in the transverse traceless gauge: 
the  coordinates of  the  test masses
are unaffected by the gravitational wave, but {\em physical} ( i.e.
{\em proper}) distances between masses are influenced by the wave.
In our case the line element is:
\bean                           \label{ds2}
\di s^2 & = & -\: \di t^2 + (\delta_{ij} + h_{ij}(t-z))\: \di x^i \:
\di x^j   =
\nonumber \\
        & = & -\: \di t^2 + \di z^2 + (1 + \Phi(t-z))\:(\di x^2 + 
                                                        \di y^2)        \; ,
\eean
where we have restricted ourselves to a purely scalar wave and we use
units $c=1$; as usual
we suppose that the wavelength of the scalar gravitational wave
is much larger than the distance between the test masses, which for
the interferometer are the two mirrors and the beam-splitter; 
we take the latter
at the origin $O$ of the coordinate system (note again that in this
gauge if the beam-splitter is at rest in the origin before the waves
comes, it will always remain there; this will not be true in the gauge
considered in the next subsection)
and so its frequency $f$
is much smaller than $1/T_0$, where $T_0$ is
the time a laser-beam takes to travel from $O$ to the mass; under this
assumption, we can consider the amplitude $\Phi(t-z)$ frozen at a
value $\Phi_0$. So, the proper distance of the mass of 
coordinates $X,Y,Z$ from the origin $O$ is:
\be             \label{L}
L = \sqrt { (1+\Phi_0) (X^2 + Y^2) + Z^2 }              \; .
\ee
The physical interpretation of (\ref{L}) is clear:
the wave acts only on the coordinates that are transverse 
with respect to its
direction of propagation.
If we denote by $L_0 \equiv \sqrt{X^2 + Y^2 + Z^2} = T_0$
the proper distance that the mass has from the origin
$O$ before the wave arrives, then from
\eq{L} we see that, at first order in $\Phi_0$, 
\be                     \label{L2}      \\
\begin{array}{rcl}
{\rm if} \; Z=0 \;\;& \Rightarrow & \;\; L = \disp{ (1 + \half \Phi_0) } L_0 \\
{\rm if} \; X=Y=0 \;\;& \Rightarrow & \;\; L = L_0      \; .
\end{array}
\ee
So, just like the ordinary gravitational
waves with  $+$ and $\times$ polarizations, the
scalar wave is  transverse,
not only in its mathematical description, eq.~(\ref{h_munu}),
but also in its physical effect.

It is now easy to compute the phase-shift produced in an
interferometer. We take an interferometer whose arms are aligned
along the $x$ and $z$ axes and we consider first 
a scalar wave traveling in the $+z$
direction. From \eq{L2} we see that the time that
the laser-beam takes in order to make $N$ round-trips
from the beam-splitter to the mirror in the $x$ direction is
\be\label{11}
T_x = (1+\half \Phi_0) \: N\: 2 L_0             \; ,
\ee
while, for the beam traveling in the $z$ direction, the travel time is
unaffected by the scalar wave,
\be\label{12}
T_z = N\: 2 L_0                 \; .
\ee
The phase-shift $\Delta \varphi$ is obtained (see e.g.
\cite{giazotto89,saulson94}) 
by multiplying the time difference $(T_x - T_z)$ by the
angular frequency $\Omega \las$ of the laser; this leads to the
result
\be             \label{delta_phi}
\Delta \varphi = \half \Phi_0 \cdot\varphi_{\rm arm}            \; ,
\ee
where 
$\varphi_{\rm arm} \equiv \Omega \las  N  2 L_0$ is the phase that
the laser-beam accumulates in $N$ round-trips.

\begin{figure}
\centering
\includegraphics[width=0.6\linewidth,angle=270]{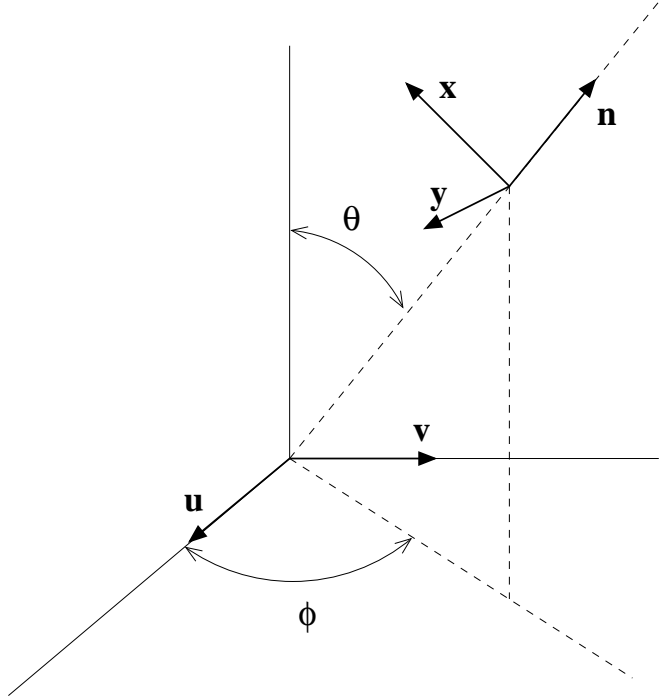}
\caption{The definitions of the versors and angles discussed in the text.}
\label{terna1}
\end{figure}

To compute the response of the interferometer for an
arbitrary direction of propagation of the wave we 
recall  that it is possible to
associate to a detector a `response tensor' $D_{ij}$ such that
the signal (the phase-shift, if the detector is an interferometer) 
induced in the detector
by a gravitational wave of polarization $e_{ij}$ is proportional
to $\tr \{ D \: e \} = D_{ij} \: e_{ji}$ \cite{For,Fla}; 
for an interferometer
whose arms are along the $\uarm$ and $\varm$ directions  the detector
tensor is
\be
D_{ij} = \hat u_i \hat u_j - \hat v_i \hat v_j\, .
\ee
The polarization tensor for a purely scalar wave
traveling along a generic direction $\n$ is (see (\ref{esmn}))
\be
e^{\rm (s)} _{ij} (\n) = \delta_{ij} - \hat n_i \hat n_j        \; ,
\ee 
and then the pattern function $F_{\rm s}(\n) \equiv D_{ij} \: e_{ji}^{\rm
(s)}$, which
describes the interaction between a scalar wave propagating along $\n$
and an interferometric detector with arms along $\uarm$ and $\varm$, is
\be             \label{F_di_n}
F_{\rm s}(\n) = - \sin^2 \theta \cos 2\phi              \; ,
\ee
where $\theta$ and $\phi$ are, respectively, the polar and azimuthal angles 
of the versor $\n$ in the reference frame in which the $x$ and $y$ axes are
co-aligned with $\uarm$ and $\varm$, (see figure \ref{terna1}).
Of course, if $\theta =0$, there is no phase shift because 
the proper length of the two arms  is modified in the same way, and it
cancels taking the difference.
However  the proper time
that the light takes to make a round trip in each of the two arms
separately is modified by the wave, and it is in principle
measurable.\footnote{VIRGO takes its output from a set of 5 photodiods
which give 5 optical lengths including the difference between the two
arms, of course, and also the common mode, i.e. the sum of the two arm
lenghts. However for the common mode the
fluctuations in the laser power do not cancel, and the sensibility is
quite limited; it is basically the same sensibility that could be obtained
measuring the common mode with two separate interferometers, one made
with one arms and the prestabilization cavity and the other with the
other arm and again the  prestabilization cavity.}
For a generic angle of incidence the cancellation does not take place
and there is a phase shift.

\begin{figure}
\centering
\includegraphics[width=0.6\linewidth,angle=0]{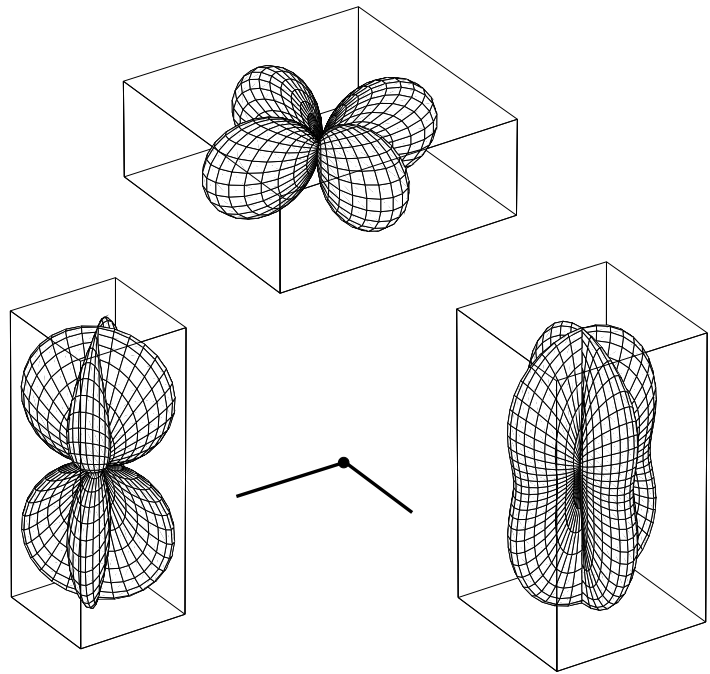}
\caption{The angular sensitivity of the interferometer for scalar
waves (top), $+$ polarization (bottom right) and $\times$ 
polarization (bottom left).
The solid lines in the middle of the figure indicate the orientation
of the arms of the interferometer.}
\label{distribuzioni}
\end{figure}

Combining (\ref{F_di_n}) with (\ref{delta_phi}) we get the phase shift
for arbitrary direction of incidence of the scalar wave~\cite{Nic}, 
\be
\Delta \varphi (\theta, \phi) = -\half \Phi_0 \cdot \varphi_{\rm arm} \cdot
                        \sin^2 \theta \cos 2\phi                \; .
\ee
The angular sensitivity of the interferometer  to scalar GWs is shown
in fig.~(\ref{distribuzioni}), together with the standard angular
pattern for the $+$ and $\times$ polarizations.

\clearpage

\subsection{Computation in the gauge $\esmn =\eta_{\mu\nu}$}
It is instructive to repeat the computation  in a different
gauge. Indeed, with a different gauge choice, we can bring the
plane wave solution of the equations of motion into the form
of \eq{h_munu} with $\eplmn$ and $\ecrmn$ unchanged, while
\be
\esmn =\eta_{\mu\nu}\, ,
\ee
so that, for a purely scalar wave, 
\be
g_{\mu\nu}=(1+\Phi )\eta_{\mu\nu}\, .
\ee
In app.~B we prove this assertion and we find the coordinate
transformation that allows to move from the previous gauge to this
one.\footnote{This gauge has been  used in ref.~\cite{SNN}. However,
the authors did not realize that in this reference frame the beam
splitter is not left at the origin by the passage of the GW, see
below, and furthermore
computed a coordinate-time interval rather than a proper-time
interval,
reaching the incorrect conclusion that the scalar GW has a longitudinal
effect, and no transverse effect. Their expression for the phase shift
and angular pattern then differs from ours. In this subsection we will
confirm the results of  section 2.1,
with a proper treatement of the scalar GW in this gauge.}

As in  section 2.1, we consider a purely scalar gravitational wave
traveling in the $+z$ direction and impinging on an interferometer whose arms are aligned
along the $x$ and $z$ axes,
and have a length $L_0$ when there is no gravitational wave. 
We will denote quantities relative to the arms by subscripts $X$ and
$Z$.
Again, we deal only with the case in which the frequency $f$ 
of the wave is much smaller then $1/T_0 = 1/ L_0$, as discussed in the 
previous section.

Consider the metric
\be
\di s^2 = (1 + \Phi) (-\di t ^2 + \di x^2 +\di y^2 +\di z^2)    \; .
                                \label{ds2_shibata}
\ee
The laser-beam follows null geodesics ($\di s^2 = 0$): so, for the
beam that propagates along the $x$ axis
\be
\frac {\di x \las}{\di t} = \pm 1 \; \Rightarrow x   \;
\las (t) = \const \pm t         \; ,            \label{x=t}
\ee
and for the beam along $z$
\be
\frac {\di z \las}{\di t} = \pm 1 \; \Rightarrow z   \;
\las (t) = \const \pm t                 \; .
\ee
The geodesic equation of motion for the mirrors and beam-splitter, in
this gauge, is~\cite{SNN}
\be                             \label{geodesic}
\left\{         \begin{array}{rcl}
x(t)   &  =   &  x_0   \\
y(t)   &  =   &  y_0   \\
z(t)   &  =   &  z_0 + \displaystyle{\half} I(t-z(t))  \\
\tau (t)   &  =   &  t + z(t)\, ,
\end{array}     \right.
\ee
where 
\be
I(t-z) \equiv \int ^{t-z} _{- \infty} \! \Phi (u) \di u         \; . 
\ee
(note that the solution for $z(t)$ is an implicit
equation). Therefore, in this gauge, the transverse coordinates are
not affected by the passage of the wave, while the longitudinal
coordinate is affected. Of course, this does not mean that the
physical effect of the wave is longitudinal. The physical effect is
determined looking at diffeomorfism invariant quantities, i.e. 
proper distances and proper times.

Since in this gauge the $x$-coordinates of
the mirrors and the beam-splitter are unaffected by the passage of the
wave,
from eq.~({\ref{x=t}}) we see  that the interval, in coordinate 
time $t$, that the laser 
takes for one round-trip in the $X$ arm is
\be
T_X = 2 L_0                     \label{tx}      \; .
\ee 
$T_X$ measures the time that the laser-beam takes in terms of the variable
$t$:  the beam leaves the
beam-splitter at $t=0$ and comes back at $t=T_X$. However, this quantity
is by definition not invariant under coordinate transformations, and
in order to deal with interference at the beam-splitter, we 
must work in terms of the beam-splitter proper time; 
it is this latter variable that measures the physical length of the arms.
Thus we call $\tau(t)$ and $z_{\rm BS} (t)$ the proper time 
and $z$-coordinate of the beam-splitter at (coordinate)
time $t$ (with initial condition $z_{\rm BS} ( - \infty) = 0$). 
From eqs.~(\ref{geodesic}),
\bean
z_{\rm BS} (t) & = &  \half \int_{-\infty}^{t-z_{\rm BS} (t)} 
                        \Phi(u) \di u   \\
\tau (t) & = & t + \half \int_{-\infty}^{t-z_{\rm BS} (t)} \Phi(u) \di u
        \; .
\eean
The proper time 
interval that the beam takes to make a round-trip in the $X$ arm is then
\bean
\tau_X  \equiv  \tau( T_X ) - \tau( 0 ) & = & T_X +
        \half \int_{-z_{\rm BS} (0)}^{T_X - z_{\rm BS} (T_X)} \Phi(u) \di u 
        \simeq          \nonumber               \\
& \simeq & T_X + \half \Phi_0 [T_X + z_{\rm BS} (0) - z_{\rm BS} (T_X)] 
        \simeq          \nonumber               \\
        & \simeq & 2 L_0 (1 + \half \Phi_0)     \; ,            \label{taux}
\eean
where we have used eq. (\ref{tx}), 
we have considered $\Phi$ ``frozen'' at $\Phi_0$ and we have
kept only the first order in $\Phi_0$. Note that 
$z_{\rm BS} (0) - z_{\rm BS} (T_X)=O(\Phi_0)$.

We now compute $T_Z$ and $\tau_Z$. Suppose that
the beam leaves the beam-splitter at $t=0$  and reaches the
mirror along the $z$ axis at $t=T_1$. Then
\be                     \label{uno}
z_{\rm BS} (0) + T_1 = Z( T_1 )         \; ,
\ee
where $Z(t)$ denotes the $z$-coordinate of the $Z$ mirror at time $t$ (with $Z( - \infty) = L_0$).
Similarly, when coming back from the mirror the beam reaches again 
the beam-splitter 
at $t = T_Z = T_1 + T_2$ 
such that 
\be                     \label{due}
Z( T_1 ) -  T_2 = z_{\rm BS} (T_Z)      \; .
\ee
Subtracting eq. (\ref{due}) from eq.(\ref{uno}) one has 
\be                     \label{tre}
T_Z = T_1 +  T_2 = 
        \left[ Z( T_1 ) - z_{\rm BS} (0)\right] +
        \left[ Z( T_1 ) - z_{\rm BS} (T_Z)\right]       \; .
\ee
The equations of motion (\ref{geodesic}) 
for the $Z$ mirror and the beam-splitter read
\bean
Z(t) & = & L_0 + \half \int _{ - \infty} ^{t - Z(t)} \Phi (u) \di u     \\
z_{\rm BS} (t) & = & \half \int _{ - \infty} ^{t - z_{\rm BS} (t)} 
                                                        \Phi (u) \di u  \; ,
\eean
which, substituted in eq. (\ref{tre}), give 
\be                                     \label{quattro}
T_Z = 2 L_0 + \half
\int _{ - z_{\rm BS} (0) } ^ { T_1 -  Z( T_1) } \Phi (u) \di u  + 
\half \int _{ T_Z - z_{\rm BS} (T_Z) } 
                        ^ { T_1 -  Z( T_1) }    \Phi (u) \di u          \; .
\ee
Because of eq.(\ref{uno}), the first integral in eq. (\ref{quattro}) is zero.
The second integral, 
 in the approximation of a frozen wave and
to first order in $\Phi_0$, gives
\bea
\half \int _{ T_Z - z_{\rm BS} (T_Z) } 
        ^ { T_1 -  Z( T_1 ) } \Phi (u) \di u & \simeq &
\half \Phi_0 \left[  T_1 -  Z( T_1 ) - T_Z 
                        + z_{\rm BS} (T_Z)   \right]    \simeq          \\
        & \simeq & \half \Phi_0 [ L_0 - L_0 - 2L_0 + 0] = -\half \Phi_0 \: 2 L_0
                \; .
\eea 
Eq. (\ref{quattro}) thus becomes
\be
T_Z = (1 - \half \Phi_0) \: 2 L_0                       \; .
\ee
Following the same steps leading to eq. (\ref{taux}), we then obtain
\be
\tau_Z \equiv \tau(T_Z) - \tau(0) \simeq T_Z (1+\half \Phi_0) 
        \simeq 2 L_0            \; .    \label{tauz}            
\ee
Eqs. (\ref{taux}) and (\ref{tauz})  agree with the result of sect.~2,
see \eqs{11}{12}, as they should. 
It is also apparent that the computation in the
gauge used in the previous section is much simpler, due to the fact
that in this case the coordinates of test masses are not affected by
the passage of the wave, and only proper distances change.

\section{Stochastic backgrounds of scalar GWs}

\subsection{General definitions}

The standard procedure described in 
\cite{Fla,All,Mag} for detection of stochastic
backgrounds of ordinary gravitational waves can be applied 
with minor modifications to the 
case of  scalar GWs.
A stochastic, gaussian and isotropic background
of scalar GWs can be characterized by the quantity
\be
\Omega_\varphi (f) \equiv \frac{1}{\rho_{\rm c}} \: \frac{\di \rho_\varphi}{\di \log f} \; ,
\ee  
where $\rho_\varphi$ is the energy density associated to the background, $f$ is the frequency and
$\rho_{\rm c}$ is the critical energy density for closing the Universe,
\be
\rho_{\rm c} = \frac{3 H_0^2}{8 \pi G}
\ee
($H_0 = 100 \: h_0 \: ({\rm km} / {\rm sec})(1 / {\rm Mpc})$ 
is the Hubble constant).
The intensity of the background is expected to be well below the
noise level of a detector: 
the detection strategy then 
consists in correlating the outputs of two (or more) detectors, 
located far enough
so that local noises as the seismic noise
are uncorrelated. One defines the quantity
\be                     \label{S}
S \equiv \int_{-T/2}^{+T/2} \! \di t \int_{-T/2}^{+T/2} 
\! \di t' \: s_1(t) \: s_2 (t') \:
                        Q(t-t')                 \; ,
\ee
where $s_a$ denotes the output of the $a$-th detector, $T$ the total
observation time and
$Q$ is a real filter function, that, for any given form of the signal, can be
determined exactly in order to maximize the signal-to-noise ratio
(SNR). We write
$s_a (t) = h_a (t) + n_a (t)$, where $h_a$ is the signal induced by the
background of scalar GWs (for an interferometer it is the phase
shift computed in the previous section),
and $n_a$ is the intrinsic noise of the $a$-th detector. 
Under the assumption of uncorrelated noises, the ensemble average (denoted by 
$\langle \cdots \rangle$) of the Fourier components of the noise satisfies
\be
\langle \: \tilde n_a ^* (f) \: \tilde n_b (f') \: \rangle =
        \delta (f-f') \: \delta_{ab} \: S_n^{(a)} (|f|)                 \; .
\ee
The functions $S_n^{(a)} (|f|)$ are known as  square spectral noise densities.
The signal-to-noise ratio is defined as
\be
\mbox{SNR} \equiv \left[ \frac{\langle S \rangle ^2}{\langle (S - \langle S \rangle)^2 \rangle}
                        \right] ^{1/4}                  \; ,
\ee
where we have used the exponent $1/4$ (instead of $1/2$) to take into account the fact that
$S$ is quadratic in the signals (see (\ref{S})).
Optimal filtering (see \cite{All,Mag}) gives
\be                     \label{snr}
\mbox{SNR} = \left[ 2T \int_0 ^{+\infty} \! \di f \: \frac{S^2 _h (f)}{S^2 _n (f)} \:
                        \Gamma ^2 (f)
                \right]^{1/4}           \; .
\ee
The various quantities that appear in the SNR are defined as follows.

\begin{itemize}
\item The function $S_h$ depends uniquely on the spectrum of the background, 
and not on the features of the detectors, and is related to
$\Omega_\varphi$ by
\be
S_h (f) = \frac{3 H_0 ^2}{4 \pi ^2  f^3} \:
\frac{1}{2+\obd} \: \Omega_\varphi (f)  \; .
\ee
Note that in the case of ordinary gravitational waves
the factor $1/(2+\obd)$ is absent. The derivation of this result is
given in app.~C. 

\item The function $S_n$ is defined as $S_n(f) \equiv \sqrt{S_n^{(1)}
(f) \: S_n^{(2)} (f)}$ and therefore
depends uniquely on the intrinsic noises of the detectors.

\item The function $\Gamma(f)$ gives a  measure of the  correlation between
the detectors, and
depends on their relative position and orientation; it is defined as
        \be                     \label{gamma1}
        \Gamma (f) \equiv \int \! \frac{\dOn}{4\pi} \: F^{(1)} _{\rm s}(\n) \: 
                F^{(2)} _{\rm s}(\n) \: e^{i \: 2\pi fd \:  \m \cdot \n}        \; ,
        \ee
where $d \cdot \m = {\bf x}^{ (1)} - {\bf x}^{ (2) }$ 
is the vector joining the detectors ($d$ is
their distance, $\m$ a versor) and the 
$F^{(a)} _{\rm s}(\n)$ are the pattern functions of
the detectors for scalar waves.

\end{itemize}

In the case of the correlation between two interferometers, it is
conventional to define the overlap reduction function $\gamma$ 
by~\cite{Chr,Fla} 
\be\label{gammaGamma}
\gamma(f)=\frac{\Gamma(f)}{F_{12}}\, .
\ee
where, for ordinary GWs
\be\label{FF12}
F_{12}\equiv
\int\frac{\dOn}{4\pi}\sum_{A=+,\times} F_A^{(1)}(\n )
F_A^{(2)}(\n )_{|\rm aligned}\, ,
\ee
and the subscript means that we must compute $F_{12}$ taking the two
interferometers to be perfectly aligned, rather than with their actual
orientation. 

This normalization 
is useful in the case of two interferometers, since in this case, for
ordinary GWs
$F_{12}=2/5$, (while for scalar GWs $F_{12}=4/15$) and it
 takes into account the reduction in sensitivity due to
the angular pattern, already present in the case of one
interferometer, and therefore $\gamma (f)$ separately
takes into account the effect
of the separation $\Delta\vec{x}$ between the interferometers, and of
their relative orientation. With this definition, $\gamma (f)=1$ if the
separation $\Delta x =0$ and if the detectors are perfectly aligned.

This normalization is instead impossible when one considers the
correlation between an interferometer and the monopole mode
of a resonant sphere, since in
this case 
$F_{12}=0$, as we will see below. Then one simply uses 
$\Gamma (f)$, which is the quantity that enters directly \eq{snr}.
Furthermore, the use of $\Gamma (f)$ is more convenient when we want
to write equations that hold independently of what detectors
(interferometers, bars, or spheres) are
used in the correlation. 
In the following we will always refer to $\Gamma (f)$  as the overlap
reduction function.

\subsection{The overlap reduction function for scalar GWs}
In this section we compute analytically the overlap reduction 
function of two 
generic detectors in the case of a background of scalar waves,
generalizing the result for ordinary gravitational waves of ref.~\cite{Fla}.

As one can see from eq.~(\ref{gamma1}), the overlap reduction function 
is obtained by averaging,
over the possible directions of propagation of the gravitational
wave, the product of the pattern functions of the two detectors, 
weighted with a phase that
depends on the delay 
in the propagation from one detector to the other. The pattern
function and detector tensors are related by
\be
F_{\rm s} (\n) = D_{ij} e_{ij}^{\rm (s)}(\n )
\ee 
As in the  case of ordinary GWs, the response tensors $D^{(a)}_{ij}$ are
normalized in such a way that the pattern functions $F_{\rm s}^{(a)}
(\n)$  take 1 as maximum
value,  varying $\n$.
 Then
we can rewrite eq.~(\ref{gamma1}) in the  form
\be\label{29}
\Gamma(f) = \duno _{ij}  \:  \G (\alpha, \m)  \:  \ddue _{kl} \label{gamma2}  \; ,
\ee
where $\alpha \equiv 2\pi f d$ and
\be       \label{Gamma}
\G (\alpha, \m)    \equiv    \int \! \frac{\dOn}{4\pi} \: 
                        e_{ij}^{\rm (s)}(\n ) \: e_{kl}^{\rm (s)}(\n ) 
                        \: e^{i \: \alpha \:  \m \cdot \n}   \; .
\ee
As we discussed in the previous section, 
the polarization tensor for a scalar wave traveling
in the direction $\n$ is
\be                             \label{en}
e_{ij}^{\rm (s)}(\n ) = \hat x_i \hat x_j + \hat y_i \hat y_j = \delta_{ij} - \hat n_i \hat n_j \; ,
\ee 
where $\x$ and $\y$ are versors perpendicular to $\n$
and to each-other (see fig. \ref{terna1}). 
The difference between the scalar case and the traditional one resides in $\G$, because
in the latter case the polarization tensors 
$e_{ij}^{(+,\times )}(\n )$ take the form:
\be     \begin{array}{rcl}
e_{ij}^{(+)}(\n ) & = & \hat x_i \hat x_j - \hat y_i \hat y_j                   \\
e_{ij}^{(\times )}(\n ) & = & \hat x_i \hat y_j + \hat y_i \hat x_j             \; .
\end{array}
\ee
We now proceed to compute the tensor $\G$. 
Note that the result we
find is absolutely general, independently
on the type of the detectors  used in the correlation, 
since the informations on the detectors 
is summarized in the response tensors $D^{(a)}_{ij}$ and not in $\G$.

Since the tensor $\G$ is symmetric in $i \leftrightarrow j$, in
$k \leftrightarrow l$, 
and in  $(i j) \leftrightarrow (k l)$ (see eq.~(\ref{Gamma})),
it can be written in the most general form as
\be             \label{Gamma2}
\G  (\alpha, \m) = A(\alpha) \A + B(\alpha) \B + C(\alpha) \C
        + D(\alpha) \D + E(\alpha) \E
\ee
where
\bean                   \label{tensori}
\A      &       =       &       \delta _{ij} \delta _{kl}       \nonumber \\
\B      &       =       &       \delta _{ik} \delta _{jl} +
                                \delta _{il} \delta _{jk}       \nonumber \\
\C      &       =       &       \delta _{ij} \hat m_k \hat m_l + 
                                \delta _{kl} \hat m_i \hat m_j  \\
\D      &       =       &       \hat m_i \hat m_j \hat m_k \hat m_l             \nonumber \\
\E      &       =       &       \delta _{ik} \hat m_j \hat m_l +
                                \delta _{jk} \hat m_i \hat m_l +
                                \delta _{jl} \hat m_i \hat m_k +
                                \delta _{il} \hat m_j \hat m_k   \nonumber
\eean
(the tensor $\epsilon _{ikm} \hat m_m \epsilon_{jln} \hat m_n +
\epsilon _{ilm} \hat m_m \epsilon_{jkn} \hat m_n$ is a linear combination of these).

To determine the coefficients $A(\alpha)$, $B(\alpha)$, $\dots$,
$E(\alpha)$, one has to contract
$\G$ (written both in the form (\ref{Gamma}) and in the form
(\ref{Gamma2}))  with the
five tensors (\ref{tensori}) and to calculate the corresponding scalar
integrals, which can be done in terms of spherical Bessel functions; one thus
obtain a  linear system with five equations, whose solution is
\be                                     \label{generale}
\left[          \begin{array}{c}
A  \\  B \\  C  \\  D  \\  E
\end{array}     \right]  (\alpha)   \:\:   =  
                \:\:  \frac {1}{\alpha^2}
\left[  \begin{array}{r}
\alpha^2 \cdot \jzero - 2 \alpha \cdot \juno + \jdue    \\
\jdue  \\
- \alpha^2 \cdot \jzero +4 \alpha \cdot \juno -5 \cdot \jdue    \\
\alpha^2 \cdot \jzero -10 \alpha \cdot \juno +35 \cdot \jdue    \\
\alpha \cdot \juno -5 \cdot \jdue   
\end{array}     \right]                         \; ,
\ee
where the $j_\ell$ are the spherical Bessel functions,
\bea
\jzero  &       \equiv  &       \frac {\sin \alpha}{\alpha}                     \\
\juno   &       \equiv  &       \frac {\jzero - \cos \alpha}{\alpha}            \\
\jdue   &       \equiv  &       3 \: \frac{\juno}{\alpha} - \jzero              \; .
\eea
The overlap reduction function for any specific two-detectors
correlation can now be obtained using this general form for $\G$ and
then plugging into \eq{29} the appropriate detector tensors. 

A particularly interesting correlation is the one between the VIRGO
interferometer and a resonant sphere.
A resonant sphere has 6 different detection
channels: 5 corresponding to the quadrupole modes with
$\ell=2$, $m=0,\pm 1,\pm 2$, and 1 corresponding to the monopole mode with
$\ell=0$, $m=0$. 
The monopole mode is especially interesting because it cannot be excited by
ordinary spin 2 GWs.
The response of a resonant
sphere to scalar GWs has been computed in ref.~\cite{BBCFL}.

\begin{figure}  
\begin{center}
\includegraphics[bb= 0 0 381 362,scale=0.50,angle=-90]{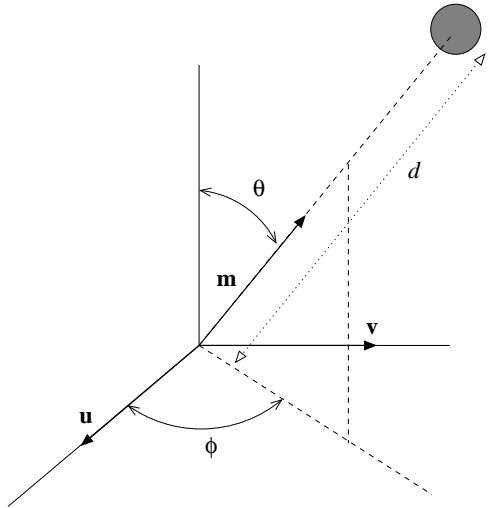}
\caption{\label{detectors} Relative position of the sphere and interferometer.}
\end{center}
\end{figure}

To correlate an interferometer,
with arms  along
$\uarm$ and $\varm$, with the monopole mode of
a resonant sphere, we use
the response tensors
\bean
\duno   &  =   &  \uarm \otimes \uarm - \varm \otimes \varm     \label{duno}\\
\ddue   &  =   &  \half \uno _3                 \; ,            \label{ddue}
\eean
and we get~\cite{Nic}
\be
\Gamma (f) = \( \sin^2 \theta \cos 2\phi \)\, j_2 (2 \pi f d)           \; , 
\ee
where $\theta$ and $\phi$ are the angular coordinates of the resonant
sphere  (i.e. of $\m$) 
with respect to the ``natural'' reference frame of the interferometer 
(see figure \ref{detectors}).

\begin{figure}
\begin{center}
\includegraphics[bb= 0 0 535 330,scale=0.50,angle=0]{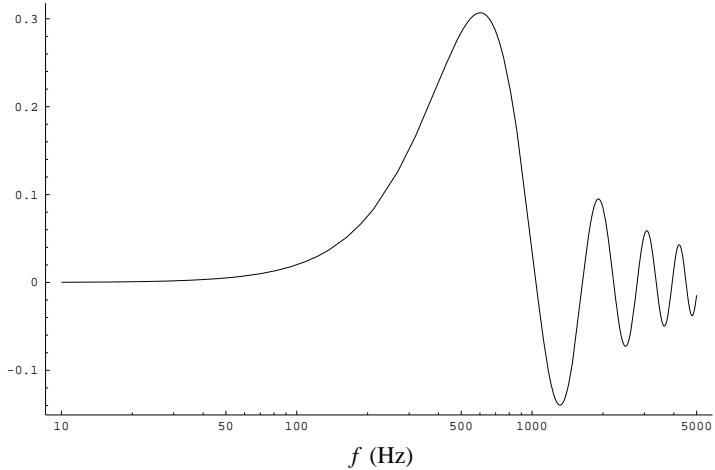}
\caption{\label{overlap_fig}
The behavior of $j_2 (2\pi f d)$ against the frequency $f$
(Hz) for
$d=270$ km, which is approximately the distance between the location of
VIRGO  and Frascati (resonant spheres).}
\end{center}
\end{figure}

Figure \ref{overlap_fig} shows the frequency dependence of this 
overlap reduction function
for a particular choice 
of the distance $d$ corresponding to the distance between the VIRGO
site and Frascati. The most striking aspect, compared to the overlap
reduction functions of interferometer-interferometer correlations, is
that the correlation vanishes if the distance between the detectors,
$d$, goes to zero. This is the opposite of what happens for 
two interferometers, where instead the correlation is maximum for
coincident detectors, so that the ideal strategy in that case
would be to have two
interferometers as close as possible, compatibly with the fact that local
noises, like the seismic noise and local  electromagnetic
disturbances should be uncorrelated (it is believed that this implies a
minimum distance of at least a few tens of kms). In this case, instead,
a sphere coincident with an interferometer makes for a totally
ineffective correlation. This is easily understood, because the
monopole mode of the sphere
has, obviously, a constant angular pattern, while an interferometer
has a dependence $\sim\cos 2\phi$ which over the solid angle 
integrates to zero. Parentetically, this shows that in this case the
function $\gamma (f)$ cannot be defined, since it should be defined
dividing $\Gamma (f)$ for the value for coincident detectors, which is
now zero. However, $\Gamma (f)$ is the quantity that enters the SNR. 

Of course, as $d\ra\infty$ the overlap reduction function again
vanishes, and therefore there is an optimal, finite value for the
product of the distance $d$ and the frequency $f$, determined by the
maximum of $j_2 (2\pi f d)$. Approximately, this maximum is reached
when
\be\label{fd}
\(\frac{f}{590{\rm Hz}}\)\(\frac{d}{270{\rm km}}\)\simeq 1\, .
\ee
We have taken the reference value 270 km, which corresponds to the
distance between the locations of VIRGO (Lat. N 43.63, Lon. E 10.50 deg)
and Frascati  (Lat. N 41.80, Lon. E 12.67 deg). Since the resonant
sphere is a narrow band detector, compared to the interferometer, this
means that the optimal correlation is obtained with a sphere that
resonate at approximately 590 Hz. We see from fig.~(\ref{overlap_fig})
that the sensitivity of the overall correlation is very strongly
affected by the value of $f$, and a value of $f\sim $ 1~kHz, while
keeping the sphere in Frascati, can easily result in the loss of
more than one order of magnitude in the SNR defined in \eq{snr}. Note
also that the minimum detectable value of $\hogw$ is quadratic with
the SNR \cite{Mag} and therefore an inappropriate value
of the resonant frquency can easily result in loosing
two or three order of magnitude  in $\hogw$.

Conversely, if for technological reasons the frequency of the resonant
sphere has to be fixed at a different value, \eq{fd} can be used to obtain
the optimal
distance between the two detectors.

\section{The interaction of the interferometer with a very light scalar}
Till now we have considered only massless scalars; it is
interesting, having in mind the extension to situations motivated by
string theory, to consider the effect of a small mass term on the
response of a detector. 

Let us first understand what it means `small'
in this context. 
We still want to treat the scalar as a
classical wave that acts coherently on the detector.
This implies first of all that $m\ll 1/L$, where $L$ is the
characteristic size of the detector. For a sphere of size of order 1
meter, this gives approximately $m< 10^{-6}$ eV, while for a km sized
interferometer $m<10^{-9}$ eV. In both cases, however, there is a
stronger limitation that comes from the fact that, if we want to
detect these scalars with a detector that works at a frequency $f_0=1$
kHz, or $\omega_0 =2\pi$ kHz, we must require that the frequency of
the massive scalar, 
$\omega_m=\sqrt{k^2+m^2}$, is of order  of $\omega_0$. Since of course
$\omega_m \geq m$, we get
\be\label{bound}
m < 2\pi\, {\rm kHz}\sim 4\times 10^{-12}\, {\rm eV}\, .
\ee
For these ultralight
scalars it still make sense to discuss their effect as a coherent
gravitational wave\footnote{Of course, in general such a ultralight
scalar would mediate an unacceptably strong fifth force: Newton's law is tested
down to length of order $\sim$~cm, so that an hypothetical fifth force could
be mediated only by a particle with mass $m > (1 \: {\rm cm})^{-1} \simeq 2 \times 
10^{-5} \:$~ eV (see e.g. ref.~\cite{TV}). Here, however, we are
considering Brans-Dicke theory, with $\obd >600$: the corrections to 
Newton's gravity, in the weak field limit, are smaller than $O(1/ \obd)$, so that even a
massless scalar is compatible with experiments.}

In the opposite limit of large mass, the single
quanta of the scalar field will behave as particles rather than waves 
and will interact incoherently with the
detector, just with separate hits 
and since they only interact gravitationally they will be totally
unobservable. 

Consider the action describing the Jordan--Brans--Dicke theory (\ref{BD}), 
with in addition a potential term for the scalar:
\bean
\s & = & \s_{\rm g} + \s_{\rm m}\, ,    \nonumber\\
\s _{\rm g} [ \gmn, \varphi ]   & = &   
                \frac {1}{16\pi} \int \! \di^4 x \sqrt{-g}
                \left[ \varphi R - \frac{\obd}{\varphi}
                \: \nabla^\mu \varphi \: \nabla_\mu \varphi 
                - V (\varphi) \right]\, ,               \nonumber       \\
\s _{\rm m} [ \psi_{\rm m}, \gmn ]      & = &   
                \int \! \di^4 x \sqrt{-g} \: \lag _{\rm m} 
                \left[ \psi _{\rm m} , \gmn \right]             \; .
\eean  
In analogy with the massless case, to obtain the field equations we vary the 
action $\s$ with respect to the metric $\Gmn$ and to the scalar
$\varphi$; then, we linearize the
equations near the background $(\etamn, \varphi_0)$, where $\varphi_0$
is a minimum of $V$.
The equations of motion in vacuum (where the matter energy-momentum tensor
$T_{\mu\nu}$ is zero) are
\bean
\R_{\mu \nu} - \half \R \: \etamn & = & - \de_\mu \de_\nu \Phi +
                        \etamn \Box \Phi                \label{massa_Rmunu} \\
\Box \Phi - m^2 \Phi & = & 0            \; ,    \label{massa_phi}
\eean
where
\begin{itemize}
\item 
$\R_{\mu \nu} , \R$ are the linearized (with respect to $\hmn \equiv (\gmn -\etamn)\:$)
        Ricci tensor and scalar curvature;  
\item 
$\Phi \equiv - (\varphi - \varphi_0) / \varphi _0$ ;
\item 
$\displaystyle{ m^2 \equiv \frac {V'' (\varphi_0) \: \varphi_0}{(3+2 \obd)} }$ .
\end{itemize}
Again in analogy to the massless case, the linearized Riemann tensor
$\R_{\mu \nu \rho \sigma}$ and the equations (\ref{massa_Rmunu}) and (\ref{massa_phi})
are invariant under gauge-transformations
\be
\left\{         \begin{array}{rclcl}
\hmn (x)        & \to & \hmn'(x)        & = &   
                                \hmn(x) - \de_{(\mu}\epsilon_{\nu)}     \\
\varphi(x)      & \to & \varphi'(x)     & = & \varphi(x)                \; ;
\end{array}     \right.
\ee 
we then define
\be
\thetamn \equiv \hmn - \etamn \left( \half h - \Phi \right)
\ee
and choose the {\em Lorentz gauge} (in analogy to electromagnetism)
\be                     \label{lorentz1}
\de _\mu \Thetamn = 0                   \; ,
\ee
by means of a transformation such that $\Box \epsilon _\nu = \de ^\mu \thetamn$.
In such a gauge the field-equations become wave-equations:
\bean
\Box \thetamn & = & 0           \label{massa_theta}\\
\Box \Phi & = & m^2 \Phi        \label{massa_phi2}              \; ,
\eean
whose solutions are plane waves (and superpositions of them):
\bean
\thetamn (x) & = & A_{\mu \nu} (\vec k) e^{i k_\alpha x^\alpha} 
                                        + {\rm c.c.}            \\
\Phi (x) & = & b (\vec k)  e^{i q_\alpha x^\alpha} + {\rm c.c.}         \; ,
\eean
where:
\be     \begin{array}{rclcl}
k^\alpha  & \equiv & ( \omega_0 ; \vec k) & \:\: &
                \omega_0 = k \equiv | \vec k |                  \\
q^\alpha  & \equiv & ( \omega_m ; \vec k) & \:\: &
                \omega_m = \sqrt{k ^2 + m^2}            \label{q_alpha}         \; ;
\end{array}
\ee
as one sees from (\ref{q_alpha}), 
the dispersion law for the modes of $\Phi$ is that
of a massive field, and the group-velocity of a wave-packet of
$\Phi$ centered in $\vec k$ is
\be
\vec v = \frac{\de \omega_m}{\de \vec k} = \frac{\vec k}{\omega_m}      \; , 
\ee
exactly the velocity of a massive particle with momentum $\vec k$.

As in the massless case, we have a residual gauge-freedom: 
we remain in the gauge
(\ref{lorentz1}) by transformations with $\Box \epsilon _\nu = 0$. 
Notice that the
Lorentz gauge is a transversality condition on the field $\thetamn$
\be
\de _\mu \Thetamn = 0   \;\; \Rightarrow \;\; k_\mu A^{\mu\nu} = 0\;
\ee
but it says nothing about the transversality of the perturbation 
$\hmn = \thetamn - \etamn \left(\displaystyle{\half} \theta - \Phi \right)$.
And, in fact, it is easy to show that, due to the field equation (\ref{massa_phi2}), it
is impossible to choose a gauge analogous to (\ref{h_munu}). In fact, 
as in the massless case discussed in app.~A,
to choose such a gauge, we should operate a transformation with:
\be             \label{NO}
\left\{         \begin{array}{rcl}
\Box \Epsm & = & 0              \\
\de_\mu \Epsm & = & -\displaystyle{\half} \theta + \Phi         \; ,
\end{array}             \right.
\ee
but now this is not compatible with the wave equations (\ref{massa_theta}),
(\ref{massa_phi2}).
Analogously, we cannot impose any linear relationship between 
$\thetamn$ and $\Phi$:
so, for example, we cannot choose a gauge in which 
$h_{\mu 0} =0$, which in the massless case
instead permits  to integrate immediately the geodesic equation.

Instead, we can choose a gauge in which a gravitational wave
propagating  in the
$+z$ direction takes the form:
\be             \label{massa_h_munu}
\hmn (t,z) = \Aplus (t-z) \: \eplmn + \Across (t-z) \: \ecrmn 
                        + \Phi (t,z) \: \etamn          \; ,
\ee
where $\eplmn$ and $\ecrmn$ are the polarization tensors
(\ref{emn}). This is similar to the gauge choice discussed in app.~B, 
but  now, because of the non-linear dispersion law, 
$\Phi$ depends {\em separately} on $t$ and $z$,
instead of depending only on their difference $t-z$.
This  makes the discussion of the physical 
effect of the wave different from the massless case.

To study the effect of the scalar wave on test-masses, we now make use of the
{\em proper reference frame} of one of these masses 
(as  the beam-splitter of 
an interferometer, for example): the analysis thus can be performed in
newtonian terms (see \cite[sect. 13.6]{MTW}), in the sense 
that the spatial coordinates $x^j$ represent 
proper distances for the observer `sitting' on the
 beam-splitter, the time variable $t$ 
represents his proper time, and the effect of the 
gravitational wave on test-masses is 
described by the equation of motion:
\be             \label{deviazione}
\frac{\di^2 x ^j}{\di t^2} =
        - {R ^j} _{0k0} \: x ^k \; ,
\ee 
where the quantities ${R ^j}_{0k0}$ are the so-called electric components
of the Riemann tensor. The problem is thus reduced to calculate
${R ^j}_{0k0}$
in the proper reference frame of this observer:  this is connected by
an infinitesimal coordinate transformation to the reference frame
where \eq{massa_h_munu} holds, and since
the linearized Riemann tensor $\R_{\mu \nu \rho \sigma}$ is
invariant under infinitesimal  gauge transformations, we can compute
it from the metric \eq{massa_h_munu}.  Restricting
to a purely scalar GW,  a straightforward computation gives
\be                     \label{massa_Ri0j0}
{\R^i}_{0j0} = \half    \left(          \begin{array}{ccc}
        - \de^2 _t      &       0               &       0       \\
        0               &       -\de^2_t        &       0       \\
        0               &       0               &       m^2
                \end{array}     \right)         _{ij}
\Phi (t,z) = - \half T_{ij} \: \de ^2 _t \: \Phi \: 
                + \: \half L_{ij} \: m^2 \Phi                   \; ,  
\ee
where $T_{ij} \equiv (\delta _{ij} - \hat n_i \hat n_j)$ is the {\em transverse
projector} with respect to the direction of propagation $\n$, and
$L_{ij} \equiv \hat n_i \hat n_j$ the {\em longitudinal projector}.
The equation of motion (\ref{deviazione}) thus becomes:
\be                             \label{massa_geo}
\frac{\di^2}{\di t^2} x_i = \left( \half \de ^2 _t \Phi \right) \cdot
        T_{ij} x_j + \left( - \half m^2 \Phi \right) 
        \cdot L_{ij} x_j                                        \; ,
\ee
from which one immediately sees what is the effect of the dilaton mass: it generates a
{\em longitudinal force} (in addition to the transverse one) proportional to $m^2$.
In the limit $m \to 0$ one recovers the treatment of the massless
case, confirming again, also from the point of view of the proper
reference frame, the results of sect.~2.

To better understand the contribution of $m$, it is convenient to
restrict to a
wave packet $\Phi(t,z)$ centered at a frequency $\omega_m =  \sqrt{ k^2 + m^2 }$; 
in this case
\be
\frac{\di^2}{\di t^2} x_i =  \left( - \half \omega _m ^2 \Phi \right)
        \cdot \left( T_{ij} + \frac{m^2}{\omega _m ^2} \: L_{ij} \right) 
        x_j                             \; .
\ee
From this equation the physical effect of the mass can be read quite
clearly. In particular
\begin{itemize}
\item[--]
for ultra-relativistic momenta ($\omega_m \simeq k \gg m$) the longitudinal components
of the force becomes negligible with respect to the transverse one;
\item[--]
in the non-relativistic limit $\omega _m \to m$ and the stress induced by the wave becomes
isotropic, as $L_{ij} + T_{ij} = \delta_{ij}$.
        \end{itemize}

Now one can compute the pattern function of a detector for these 
massive scalar waves. In a generic gauge the response of the detector
is obtained contracting the 
detector tensor $D_{ij}$ with the 
tensor ${R^i}_{0j0}$, which in the massive case is proportional to
$\omega _m ^2 T_{ij} + m^2 L_{ij}$.
Note that the pattern function of an interferometer for 
detection of such massive scalar waves
is identical to the one computed in the massless case: 
in fact, the contraction of $\omega _m ^2 T_{ij} + m^2 L_{ij}$
with $D_{ij}$ is the sum of a term containing $D_{ij} T_{ij}$ and a 
term containing $D_{ij} L_{ij}$;
both these terms are proportional to 
$D_{ij} \: \hat n_i \hat n_j$, since $D_{ij}$ is traceless, for 
an interferometer, and, so the only dependence on the mass $m$ is 
in the overall factor  of the signal and is independent
of $\n$. In particular, 
for $\omega_m \to m$ the signal goes to zero. 
The pattern function is instead mass-dependent in the case of a
detector with a non-traceless
response tensor, as for the monopole mode of the sphere, or for
the common mode of an interferometer, for which 
$D_{ij} = \hat u_i \hat u_j + \hat v_i \hat v_j$.   
\section{The string dilaton and moduli fields}
It is clearly important to understand whether these results, that have
been obtained in the context of Brans-Dicke theory,
can be applied to the physically more interesting case
of the dilaton and the other scalar fields predicted by string
theory. 

While the dilaton-graviton sector of the low energy sector of string
theory is the same as a Brans-Dicke theory, 
the situation is quite different
for the interaction of the dilaton with matter. 
Since in the string case $\obd =-1$, the dilaton is coupled with a
strength of the same order as the graviton, and produces unacceptable
deviations from general relativity, unless a non-zero dilaton mass is
generated see e.g.~\cite{TV}. 
The  radius of the
non-universal force that it mediates must be smaller than about 1~cm, or 
$m >2\times 10^{-5}$ eV. Therefore the analysis of the previous
section, which was valid for $m<4\times 10^{-12}$ eV, does not apply to a
massive dilaton.  Indeed, 
it is also in general not easy to reconcile such light scalar
particles with cosmology, see e.g.~\cite{CCQR, BKN}, although there are
mechanisms that solve the cosmological problems created by light 
scalars, typically introducing a second short  stage of inflation
that dilutes the dilaton overproduced by oscillations around their
quadratic potential.
 
Actually, there is the possibility to circumvent this bound on the
mass, with a mechanism that has been proposed 
by Damour and Polyakov \cite{DP}.
Assuming some
form of universality in the string loop corrections, it is possible to
stabilize a massless 
dilaton during the cosmological evolution, at a value where
it is essentially decoupled from the matter sector. In this case,
however, the dilaton becomes  decoupled also from the detector, 
since the dimensionless coupling of the dilaton to matter 
($\alpha$ in the notation of~\cite{DP}) is smaller
than $10^{-7}$ (see also~\cite{DE}). Such a dilaton
would then be  unobservable at VIRGO, although it
could still produce a number of small deviations from General
Relativity which might in principle be observable improving by several
orders of magnitude the experimental tests of the equivalence
principle~\cite{Dam}. 

So, in both cases, the analysis done for Brans-Dicke theory does not appear
to be relevant for string theory. However, it is clear that our
present understanding of the string dilaton and moduli is incomplete
and presents a number of unsettled issues, including the
non-perturbative mechanism for mass generation, or the stabilization at
the minimum of the potential~\cite{BS}, 
and a definite conclusion is probably premature.

\vspace{5mm}

{\bf Acknowledgments} We thank Danilo Babusci, Stefano Braccini,
Maura Brunetti, 
Ramy Brustein, Francesco Fucito and Thibault Damour for useful discussions.

\begin{appendix}
\section{Gravitational waves in the Jordan--Brans--Dicke theory}
The Jordan--Brans--Dicke theory is described in the {\em Jordan--Fierz frame} by the action
\bean
\s & = & \s _{\rm g} \left[ g_{\mu \nu} , \varphi \right]
                + \s _{\rm m} \left[ \psi _{\rm m} , g_{\mu \nu} \right]        \\
\s _{\rm g}     & \equiv &      \frac {1}{16\pi} \int \! \di^4 x \sqrt{-g}
                \left[ \varphi R - \frac {\obd}{\varphi}
                \nabla^\mu \varphi \nabla_\mu \varphi \right]   
        \label{jbd}             \\
\s _{\rm m}     & \equiv &      \int \! \di^4 x \sqrt{-g} \: \lag _{\rm m} 
                \left[ \psi _{\rm m} , g_{\mu \nu} \right]  \; ;
\eean
where:  \begin{itemize}
\item
$g_{\mu \nu}$  is the metric tensor, with which one constructs all the covariant quantities, such
as the scalar curvature $R$, covariant derivatives $\nabla$, etc.;
\item
$\varphi$ is a scalar field;
\item
$\obd$ is a parameter;
\item
$\psi _{\rm m}$ are the matter fields, such as fermions, gauge fields, etc.;
\item
$\lag _{\rm m}$ is their lagrangian.
        \end{itemize}
To obtain the field equation one as to vary the action $\s$ 
with respect to $\Gmn$ and 
to $\varphi$; this leads to 
\bean
R_{\mu \nu} - \half R \: \gmn  = \frac{8\pi}{\varphi} \: T_{\mu \nu} 
        \!\!\! & + & \!\!\! \frac{\obd}{\varphi^2} \left[ \de_\mu \varphi \: \de_\nu \varphi
        - \half \gmn g^{\alpha \beta} \: \de_\alpha \varphi \: 
        \de_\beta \varphi \right] +             \nonumber       \\ 
        & & +\frac{1}{\varphi}
        \left[ \nabla_\mu \de_\nu \varphi - \gmn g^{\alpha \beta} \:
        \nabla_\alpha \de_\beta \varphi \right]                 \label{massless_Rmunu}  \\
R - \frac{\obd}{\varphi^2} \: \Gmn \: \de_\mu \varphi \: \de_\nu \varphi
        \!\!\! & + & \!\!\! 2 \frac{\obd}{\varphi} \: \Gmn \: \nabla_\mu \de_\nu \varphi
        =  0                            \label{massless_phi}            \; ,
\eean
where
\be             \label{tmunu}
T_{\mu \nu} \equiv - \frac{2}{ \sqrt{-g} } 
        \frac{\delta \s _{\rm m}}{\delta \Gmn}
\ee
is the matter-fields energy-momentum tensor.

Eq. (\ref{massless_Rmunu}), multiplied by $\Gmn$, gives
\be
-R = \frac{8\pi}{\varphi} \: T 
        - \frac{\obd}{\varphi^2} \: \de_\mu \varphi \: \de^\mu \varphi
        - \frac{3}{\varphi} \: \nabla_\mu \de^\mu \varphi
\ee
that, substituted in eq. (\ref{massless_phi}), leads to:
\be             \label{massless_phi2}
\frac{ (3 + 2 \obd) }{\varphi} \: \nabla_\mu \de^\mu \varphi 
        = \frac{8\pi}{\varphi} \: T             \; .
\ee 
Eqs.~(\ref{massless_Rmunu}) and (\ref{massless_phi2}) are our basic
field equations.
We now study small perturbations around a background configuration: 
we choose as background
the Minkowski metric and $\varphi (x) = \varphi_0 = 
\displaystyle{\frac{4+2\obd}{(3+2\obd)G}} \:$, 
in order to have the correct post-Newtonian limit and to obtain
General Relativity when 
$\obd \to \infty \:$, as showed in \cite{Wil}; we consider
\be
\left\{         \begin{array}{rcl}
g_{\mu \nu}(x)  & = &   \etamn + \hmn (x)               \\
\varphi (x)     & = &   \varphi_0 + \delta\varphi (x)           \; ,
\end{array}     \right.
\ee  
with $| \hmn | \ll 1$ and $| \delta\varphi | \ll \varphi _0$.
We call $\R_{\mu \nu \rho \sigma}$, $\R_{\mu \nu}$ and $\R$ \label{linearizza}
the linearization to first order in $\hmn$ of the
corresponding quantities $R_{\mu \nu \rho \sigma}$, $R_{\mu \nu}$ and $R$;
one has \cite{MTW}
\be                     \label{lin_Rmnrs}
\R_{\mu \nu \rho \sigma} = \half \left\{
                \de_\mu \de_\beta h_{\alpha \nu}
        +       \de_\nu \de_\alpha h_{\mu \beta}
        -       \de_\alpha \de_\beta h_{\mu \nu}
        -       \de_\mu \de_\nu h_{\alpha \beta}
        \right\}        \; .
\ee
The linearization of
the field-equations (\ref{massless_Rmunu}) 
and (\ref{massless_phi2})  in vacuum ($T_{\mu \nu} = 0$) gives
\bean
\R_{\mu \nu} - \half \etamn \R & = & - \de_\mu \de_\nu \Phi + 
                \etamn \Box \Phi                \label{lin_Rmunu}       \\
\Box \Phi & = & 0                       \label{lin_Phi}         \; ,
\eean
with $\Phi(x) \equiv - \delta \varphi(x) / \varphi_0$.
In analogy to General Relativity, we can define a transformation acting on our fields
that leave unchanged the linearized Riemann tensor ${\R^\mu}_{\nu\rho\sigma}$ (and consequently
equations (\ref{lin_Rmunu}) and (\ref{lin_Phi})): such an {\em infinitesimal gauge 
transformation} with parameter $\Epsm$ is
\be                     \label{gauge}
\left\{         \begin{array}{rclcl}
\hmn (x)        & \to & \hmn'(x)        & \equiv &      
                \hmn(x) - \de_{(\mu}\epsilon_{\nu)}     \\
\Phi (x)                & \to & \Phi' (x)       & \equiv &      \Phi (x)        \; .
                \end{array}             
\right.
\ee
It is straightforward to verify, by direct substitution in
(\ref{lin_Rmnrs}),  that 
${\R^\mu}_{\nu\rho\sigma}$ is gauge-invariant.
We want to use this gauge
freedom to obtain a wave-equation.
We thus define
\bean                   \label{def_theta}
\thetamn & \equiv &  \hmn -\half \etamn h + \etamn \Phi \\
\theta   & \equiv &  \Etamn \thetamn = - h + 4 \Phi             \; ,
\eean
with $h \equiv \Etamn \hmn$; the transformation that 
expresses    $\hmn$ in terms of  $\thetamn$ has the
same form
\bean                   \label{def_h}
\hmn & = &  \thetamn -\half \etamn \theta + \etamn \Phi \\
h    & = &  - \theta + 4 \Phi           \; .
\eean
Substituting (\ref{def_h}) into (\ref{lin_Rmunu}) one obtains the field 
equation for $\thetamn$
\be             \label{eq_thetamn}
\Box \thetamn - \de_\mu \left( \de^\alpha \theta_{\alpha \nu}\right) -  
                \de_\nu \left( \de^\alpha \theta_{\alpha \mu}\right) +
        \etamn \de^\beta \left( \de^\alpha \theta_{\alpha \beta}\right)
                = 0                     \; .
\ee
Inserting eq.~(\ref{gauge}) into
eq.~(\ref{def_theta}) one finds immediately that under gauge-transformations
\be             \label{gauge_theta}
\left\{         \begin{array}{rcrcl}
\thetamn (x)    & \to & \thetamn'(x)    & = &   
        \thetamn(x) - \de_{(\mu}\epsilon_{\nu)} + \etamn 
                                \de^\alpha \epsilon_\alpha      \\
\theta(x)       & \to & \theta'(x)      & = &
                \theta(x) + 2 \de^\alpha \epsilon_\alpha        \\
\Phi(x)         & \to & \Phi'(x)        & = &    \Phi(x)                \; ;
\end{array}                                     \right.
\ee   
By choosing $\Epsm$ such that $\Box \epsilon _\nu = \de ^\mu \thetamn$, one has
\be             \label{lorentz}
\de ^\mu \thetamn' = 0          \; ,
\ee
and, so, a wave equation for $\thetamn'$ (see eq. (\ref{eq_thetamn}))
\be
\Box \thetamn'  =  0    \, .            \label{campo_theta}
\ee
In this gauge (we will call it {\em Lorentz gauge}, 
in analogy to electromagnetism) the solutions
are plane waves and their superpositions (we omit the $(\:\:)'$)
\bean
\thetamn (x) & = & A_{\mu \nu} ( \vec k ) \: 
        e^{i\: k^\alpha x_\alpha} \: + \:{\rm c.c.}     \label{onda_piana} \\
\Phi (x) & = & b ( \vec k ) \: e^{i\: k^\alpha x_\alpha} 
                                \: + \: {\rm c.c.}      \; ,    
\eean
with the following conditions (deriving from the field-equations and from (\ref{lorentz}))
\bean
k^\alpha k_\alpha &  = & 0              \\
k^\mu A_{\mu \nu} & = & 0       \label{trasverso}       \; ;
\eean
the latter is a transversality condition for $A_{\mu \nu}$.

Once we have chosen the Lorentz gauge, we can still operate transformations with
$\Box \Epsm = 0$; we thus take $\Epsm$ such that
\be
\left\{         \begin{array}{rcl}
\Box \Epsm & = & 0              \\
\de_\mu \Epsm & = & \displaystyle{ -\half \theta + \Phi}                
\end{array}             \right.
\ee
(it is possible bacause in vacuum in Lorentz gauge $\Box \theta = \Box \Phi =0$);
\label{massless}                \\
one has (see (\ref{gauge_theta}) and (\ref{def_h}))
\be             \label{theta_2Phi}
\theta = 2 \Phi \; \Rightarrow \;
        \hmn = \thetamn                 \; ;
\ee
that means that $\hmn$ too is a {\em plane transverse wave}.

Again, we have not yet completely fixed the gauge: we satisfy our conditions
\be             \label{vincoli}
\left\{         \begin{array}{ccc}
\de^\mu \thetamn & = & 0                \\
\theta          & = & 2 \Phi            
\end{array}             \right.
\ee
also by operating gauge transformations with
\be
\left\{         \begin{array}{rcl}
\Box \Epsm & = & 0              \\
\de_\mu \Epsm & = & 0           \; .
\end{array}             \right.
\ee
Consider the case in which the wave is propagating in $+z$ direction: then
\bean
k^\mu = (k , 0 , 0 , k) &&\\
k^\mu A_{\mu \nu} = 0   & \: \Rightarrow \: & A_{0 \nu} = - A_{3 \nu}   \\
                                          & & A_{\nu 0} = - A_{\nu 3}   \\
                                & & A_{0 0} = - A_{3 0} = + A_{3 3}     \; .
\eean
Let us make a degrees-of-freedom counting for $A_{\mu \nu}$:
\begin{itemize}  
\item[--]
we started with 10 ($=$ independent components of a symmetric tensor);
\item[--]
transversality leads to 7: it ``kills'' only 3 (instead of 4) because of symmetry of $A_{\mu\nu}$;
\item[--]
the condition $\theta = 2 \Phi$ leads to 6: we choose $A_{0 0}$,
$A_{1 1}$, $A_{2 2}$, $A_{2 1}$, $A_{3 1}$, and $A_{3 2}$ as independent components;
\item[--]
further gauge freedom permits us to put to zero 3 of the 6 components
(three rather than four, because of condition $\de_\mu \Epsm = 0$).
\end{itemize}
Thus taking
\be
\left\{         \begin{array}{rcl}
\epsm (x) & = & \tilde \epsm (\vec k) \: e^{i\: k^\alpha x_\alpha} 
                                \: + \: {\rm c.c.}              \\
k^\mu \tilde \epsm & = & 0
                \end{array}             \right.
\ee
the action of the gauge-transformation on $A_{\mu \nu}$ is
(see (\ref{gauge_theta}) and (\ref{onda_piana}))
\be
A_{\mu \nu}  \to  A_{\mu \nu} ' = A_{\mu \nu} - i k_{(\mu} 
                                                \tilde \epsilon_{\nu)}  \; ;
\ee
or, for the 6 components we are interested in,
\be     \label{gauge_A}         \begin{array}{rcl}
A_{0 0} &       \to     &       A_{0 0} + 2 \: i k \tilde \epsilon_0    \\
A_{1 1} &       \to     &       A_{1 1} \\
A_{2 2} &       \to     &       A_{2 2} \\
A_{2 1} &       \to     &       A_{2 1} \\
A_{3 1} &       \to     &       A_{3 1} - i k \tilde \epsilon_1 \\
A_{3 2} &       \to     &       A_{3 2} - i k \tilde \epsilon_2 \; .
\end{array}
\ee 
Notice that $ A_{1 1}$, $A_{2 2}$ and $A_{2 1} = A_{1 2}$ are invariant: we thus choose
$\tilde \epsilon_0$, $\tilde \epsilon_1$, $\tilde \epsilon_2$ in order to ``kill'' the others.
\label{elimina}
We have now completely fixed the gauge.

Does $\hmn$ depend on the field $\Phi$? Eq. (\ref{theta_2Phi}) tells us
$h = h_{1 1} + h_{2 2} =  2 \Phi$.\\
Summarizing, we can say that, {\em in this gauge}, the metric perturbation $\hmn$ 
produced by a gravitational wave propagating in the $+z$ direction takes the form
\be                             \label{hmunu}
\hmn (t-z) =
        A ^{(+)} (t-z)\: \eplmn + A ^{(\times)} (t-z)\: \ecrmn + \Phi (t-z)\: 
        \esmn               \; ,
\ee
where $e_{\mu\nu}^{+,\times , \, {\rm s}}$ are given in \eqs{emn}{esmn}.
\section{Equivalence of the two gauges}
In this appendix we show that it is possible to choose a gauge (which
we denote writing  a tilde  over quantities
evaluated in this gauge) in which 
the metric perturbation $\tilde h_{\mu\nu}$ 
has the form (\ref{hmunu}), with
$\tilde A^{(+)}          =       A^{(+)},
\tilde A^{(\times)}      =       A^{(\times)},  
\tilde \Phi              =       \Phi ,
\tilde \eplmn            =      \eplmn  ,       
\tilde \ecrmn            =      \ecrmn  ,       $
but
\be
\tilde \esmn     = {\rm diag} (-1,+1,+1,+1) =\etamn             \; .
\ee
The quantities without a $\tilde {(\:\:)}$ refer to the gauge used
in the previous appendix and in section 2.1. 
We call ``transverse'' the former gauge, and 
``conformal'' the latter: we want to find
a gauge transformation that passes from the transverse 
gauge to the conformal one.
In order to have a wave equation for $\thetamn$ one has to keep the
Lorentz gauge condition: this condition is mantained by imposing
$\Box \epsm = 0$. By choosing
\be
\left\{  \begin{array}{rcl}  
\de_\mu \Epsm   &  =  & - \displaystyle{ \half \theta }  \\
\Box \Epsm   &  =   &  0
\end{array} \right.
\ee
which is possible because in Lorentz gauge in vacuum $\Box \theta = 0$,
we obtain a traceless $\thetamn'$ and so (see eq. (\ref{def_h}))
\be
\hmn' = \thetamn' + \Phi \: \etamn              \; .
\ee  
In analogy with appendix A, acting  with a
transformations with
\be
\left\{  \begin{array}{rcl}
\de_\mu \Epsm   &  =  &  0   \\
\Box \Epsm   &  =  &  0    
\end{array} \right.
\ee
we can now eliminate the appropriate components of $\thetamn '$
(see equations (\ref{gauge_A})), in order to obtain exactly
$\tilde \hmn$. It is then easy to prove that $\tilde A^{(+)} = A^{(+)}$ and
$\tilde A^{(\times)} = A^{(\times)}$, by examining the action of
gauge transformations on those amplitudes.

The two gauges are thus equivalent: and in fact it is easy to exibit the
coordinate transformation that relates them (the
coordinates with a $(\:\:) '$ denote the transverse gauge and we now limit 
to the case of a purely scalar wave)
\be                             \label{primate}
\left\{    \begin{array}{rcl}  
x'  &  =  &  x  \\
y'  &  =  &  y  \\
z'  &  =  &  z - \displaystyle{\half I(t-z)}  \\
t'  &  =  &  t + \displaystyle{\half I(t-z)}            \; ,    
\end{array}   \right.
\ee     
where
\be
I(t-z) \equiv \int ^{t-z} _{- \infty} \! \Phi (u) \di u         \; . 
\ee
In fact
\be  
\left\{    \begin{array}{rcl}  
\di x'  &  =  &  \di x  \\
\di y'  &  =  &  \di y  \\
\di z'  &  =  &  \left( 1 + \displaystyle{\half} \Phi \right) \di z - 
                \displaystyle{\half} \Phi \: \di t  \\
\di t'  &  =  &  \left( 1 + \displaystyle{\half} \Phi \right) \di t - 
                \displaystyle{\half} \Phi \: \di z      \; ,
\end{array}   \right.
\ee
and
\be
\di s^2  =  (1 + \Phi) (-\di t^2 + \di x^2 + \di y^2 +\di z^2) =
 \etamn \: \di {x'}^\mu \di {x'}^\nu    
        + \Phi \: (\di {x'}^2 + \di {y'}^2)     \, .
\ee
As we have seen in section  2.1, 
the physical meaning of the primed coordinates
is that they are comoving 
with free-falling test-masses, initially at
rest, and $t'$ is proper time. 
We can check this assertion in the conformal gauge, by making use of 
the solution of the  geodesic equations of motion found in \cite{SNN},
see \eq{geodesic}.
By substituting \eq{geodesic} into the  
transformations (\ref{primate}) one finds
\be
\left\{         \begin{array}{rcl}
x'(t)   &  =   &  x_0   \\
y'(t)   &  =   &  y_0   \\
z'(t)   &  =   &  z_0   \\
t'(t)   &  =   &  t + z(t) - z_0 = \tau (t) + \const    \; ,
\end{array}     \right. 
\ee
that proves that, in the primed coordinates, bodies initially at rest remain
at rest.
\section{Relationship between $S_h (f)$ and $\Omega_\varphi (f)$}
When dealing with stochastic backgrounds of ordinary 
GWs one defines $S_h (f)$ as follows~\cite{Mag}. One expands
the metric perturbation in plane waves
\be                     \label{espandi_hij}
h_{ij} (t, \vec x) = \sum_{A = +, \times}\int_{-\infty}^{+\infty} \! \di f
                \int \! \di \Omega_\n \: h_A (f, \n) \:
                e^{2\pi i \: f(t - \vec x \cdot \n)} \:e^A _{ij} (\n)   \; , 
\ee
where the  ensemble average of the Fourier modes is 
\be                     \label{media}
\langle \: h_A ^* (f, \n) \: h_{A'} (f', \n') \: \rangle =
        \delta (f-f') \: \frac{1}{4\pi} \: \delta^2 (\n, \n') \: \delta_{A A'} 
        \: \frac{1}{2} \: S_h (f)    \; ,
\ee
and $\delta^2 (\n, \n') \equiv \delta(\phi - \phi') 
\: \delta(\cos \theta - \cos \theta ') \:$.
By inserting eq. (\ref{espandi_hij}) in the expression 
$\rho_{\rm gw} = 1/(32 \pi G) \: \langle \: \dot h_{ij} \dot h_{ij} \: \rangle$ for 
the energy density of the background, one obtains
\be
\rho_{\rm gw} = \frac{4}{32\pi G} \int_{f=0} ^{f=+\infty} \! \di (\log f) \: f \: (2\pi f)^2
                \: S_h (f)              \; ,
\ee
so that
\be
S_h (f) = \frac{3 H_0 ^2}{4 \pi ^2  f^3} \: \Omega_\varphi (f)          \; .
\ee
In the case of scalar waves, we write
\be                     \label{espandi_hij_s}
h^{\rm (s)} _{ij} (t, \vec x) = \int_{-\infty}^{+\infty} \! \di f
                \int \! \di \Omega_\n \: h_{\rm (s)} (f, \n) \:
                e^{2\pi i \: f(t - \vec x \cdot \n)} \:e^{\rm (s)} _{ij} (\n)   \; , 
\ee
and
\be                     \label{media_s}
\langle \: h _{\rm (s)} ^* (f, \n) \: h _{\rm (s)} (f', \n') \: \rangle =
        \delta (f-f') \: \frac{1}{4\pi} \: \delta^2 (\n, \n') \: \frac{1}{2} \: S_h (f)    \; .
\ee
We want to relate this new function $S_h (f)$ to the energy density $\rho_\varphi$ of the field $\varphi$.
It is convenient to rewrite the field-equation of the Jordan--Brans--Dicke  theory in the 
Einstein frame, that it is related to the Jordan--Fierz one by the conformal transformation
\be
\gmn ^{\rm E} \equiv \(\frac{\varphi}{\varphi_0}\) \gmn \; .
\ee
In this frame, the field-equation for the metric has the standard General Relativity form
\be
R_{\mu \nu} ^{\rm E} - \half \gmn ^{\rm E} \: R ^{\rm E} =
        \frac{8\pi}{\varphi_0}\: T_{\mu \nu} ^{\rm E}
        + \frac{3+2\obd}{2 \varphi^2} \left[
        \de_\mu \varphi \: \de_\nu \varphi - \half \gmn ^{\rm E}
        \: g^{\alpha \beta} _{\rm E} \: \de_\alpha \varphi \:
        \de_\beta \varphi
        \right]                 \; ,
\ee
and the energy-momentum conservation law becomes
\bean
\nabla _{\rm E}^\nu \left[
        T ^{\rm E} _{\mu \nu} + \frac{(3+2\obd) \varphi_0}{16 \pi \: \varphi^2}
        \left( \de_\mu \varphi \: \de_\nu \varphi - \half \gmn ^{\rm E}
        \: g^{\alpha \beta} _{\rm E} \: \de_\alpha \varphi \:
        \de_\beta \varphi
        \right) \right]  & \equiv &                     \nonumber \\
        \equiv \nabla _{\rm E} ^\nu \left[
        T ^{\rm E} _{\mu \nu}  + T ^{(\varphi)} _{\mu \nu}
        \right]  =  0           \; ,    & &                     
\eean
so we define $T ^{(\varphi)} _{\mu \nu}$ as the energy-momentum tensor of the field $\varphi$.
To first order in $\hmn ^{\rm E} \equiv (\gmn ^{\rm E} - \etamn)$ and 
$\Phi \equiv \displaystyle{- \left(\frac{\varphi - \varphi_0}{\varphi_0}\right)}$
we have
\be
T ^{(\varphi)} _{\mu \nu} = \frac{(3+2\obd) \varphi_0}{16 \pi}
        \left( \de_\mu \Phi \: \de_\nu \Phi - \half \etamn
        \: \de^\alpha \Phi \: \de_\alpha \Phi
        \right)         \; ;
\ee
the energy density is
\be
\rho_\varphi \equiv T ^{(\varphi)} _{00} = \frac{(3+2\obd) \varphi_0}{32 \pi}
        \left[ \dot \Phi ^2 + \left( \vec \nabla \Phi   \right)^2 \right]               \; ,
\ee
that, by using field equation $\Box \Phi = 0$ and averaging over
several wave-lengths, reduces to
\be
\rho_\varphi = \frac{(4+2\obd)}{32 \pi G} \langle \:
        2 \dot \Phi ^2 \: \rangle               \; ,
\ee
where we used $\varphi_0 = (4+2\obd)/G(3+2\obd)$, as discussed in app.~A.
As $h_{ij}^{\rm (s)} = \Phi \:  e_{ij}^{\rm (s)}$, using eqs. (\ref{espandi_hij_s}) and 
(\ref{media_s}) we have
\bean
\rho_\varphi & = & \frac{(4+2\obd)}{32 \pi G} \langle \:
        \dot h^{\rm (s)} _{ij} \dot h^{\rm (s)} _{ij} \: \rangle  =             \\
& = & \frac{2(4+2\obd)}{32\pi G} \int_{f=0} ^{f=+\infty} \! \di (\log f) \: f \: (2\pi f)^2
                \: S_h (f)              \; ,
\eean
so that
\be
S_h (f) = \frac{3 H_0 ^2}{4 \pi ^2  f^3} \:
\frac{1}{2+\obd} \: \Omega_\varphi (f)  \; .
\ee
\end{appendix}

\end{document}